\documentclass[10pt, conference]{IEEEtran}
\IEEEoverridecommandlockouts
\usepackage{cite}
\usepackage{amsmath,amssymb,amsfonts}
\usepackage{bbm}
\usepackage{algorithmic}
\usepackage{graphicx}
\usepackage{textcomp}
\usepackage{xcolor}
\usepackage{soul}
\usepackage{siunitx}
\usepackage{subcaption}
\usepackage[printonlyused]{acronym}
\usepackage{tikz}
\usepackage{textcomp}
\usepackage{hyperref}

\acrodef{mmW}{millimeter-wave}
\acrodef{sub-THz}{sub-terahertz}
\acrodef{CS}{compressive sensing}
\acrodef{CPR}{compressive phase retrieval}
\acrodef{BS}{base station}
\acrodef{UE}{User equipment}
\acrodef{Tx}{transmitter}
\acrodef{Rx}{receiver}
\acrodef{AWV}{antenna weight vector}
\acrodef{LOS}{line-of-sight}
\acrodef{NLOS}{non-line-of-sight}
\acrodef{BA}{beam alignment}
\acrodef{IA}{initial access}
\acrodef{AoA}{angles-of-arrival}
\acrodef{AoD}{angle-of-departure}
\acrodef{RSS}{received signal strength}
\acrodef{TG}{Terragraph}
\acrodef{MP}{matching pursuit}
\acrodef{ML}{machine learning}
\acrodef{PN}{pseudorandom noise}
\acrodef{QPD}{quadratic phase distribution}
\acrodef{SA}{sub-array}
\acrodef{CNN}{convolutional neural network}
\acrodef{FCnet}{fully-connected neural network}
\acrodef{MLP}{multi-layer perceptron}

\newcommand\copyrighttext{%
  \footnotesize \textcopyright 2021 IEEE. Personal use of this material is permitted.
  Permission from IEEE must be obtained for all other uses, in any current or future
  media, including reprinting/republishing this material for advertising or promotional
  purposes, creating new collective works, for resale or redistribution to servers or
  lists, or reuse of any copyrighted component of this work in other works. 
}
\newcommand\copyrightnotice{%
\begin{tikzpicture}[remember picture,overlay]
\node[anchor=south,yshift=10pt] at (current page.south) {\fbox{\parbox{\dimexpr\textwidth-\fboxsep-\fboxrule\relax}{\copyrighttext}}};
\end{tikzpicture}%
}

\DeclareMathOperator*{\argmax}{arg\,max}

\def\BibTeX{{\rm B\kern-.05em{\sc i\kern-.025em b}\kern-.08em
    T\kern-.1667em\lower.7ex\hbox{E}\kern-.125emX}}

\begin{document}
\bstctlcite{IEEEexample:BSTcontrol}


\title{Machine Learning Assisted Phase-less Millimeter-\\Wave Beam Alignment in Multipath Channels\\

\thanks{This work is supported by NSF under grant 1718742. This work was also supported in part by the ComSenTer and CONIX Research Centers, two of six centers in JUMP, a Semiconductor Research Corporation (SRC) program sponsored by DARPA.}
}


\author{\IEEEauthorblockN{Benjamin W. Domae, Ruifu Li, and Danijela Cabric	}
		
\IEEEauthorblockA{\textit{Electrical and Computer Engineering Department,} \\
\textit{University of California, Los Angeles}\\
bdomae@ucla.edu, doanr37@ucla.edu, danijela@ee.ucla.edu }
}


\maketitle
\copyrightnotice

\begin{abstract}
Communication systems at \ac{mmW} and sub-terahertz frequencies are of increasing interest for future high-data rate networks. One critical challenge faced by phased array systems at these high frequencies is the efficiency of the initial beam alignment, typically using only phase-less power measurements due to high frequency oscillator phase noise. Traditional methods for beam alignment require exhaustive sweeps of all possible beam directions, thus scale communications overhead linearly with antenna array size. For better scaling with the large arrays required at high \ac{mmW} bands, compressive sensing methods have been proposed as their overhead scales logarithmically with the array size. However, algorithms utilizing machine learning have shown more efficient and more accurate alignment when using real hardware due to array impairments. Additionally, few existing phase-less beam alignment algorithms have been tested over varied secondary path strength in multipath channels. In this work, we introduce a novel, machine learning based algorithm for beam alignment in multipath environments using only phase-less received power measurements. We consider the impacts of phased array sounding beam design and machine learning architectures on beam alignment performance and validate our findings experimentally using 60 GHz radios with 36-element phased arrays. Using experimental data in multipath channels, our proposed algorithm demonstrates an 88\% reduction in beam alignment overhead compared to an exhaustive search and at least a 62\% reduction in overhead compared to existing compressive methods.
\end{abstract}


\section{Introduction}
Due to large swaths of available bandwidth, upper \ac{mmW} and \ac{sub-THz} are strong candidates for future high data rate cellular and wireless local area networks. To overcome the significantly higher path loss in these bands, \ac{mmW} wireless systems must use strongly directional antenna patterns for data communication. \ac{UE} typical requires phased array antennas with many elements to achieve electronically steerable directional beams. Additionally, during an initial connection to a \ac{BS}, \ac{IA}, the \ac{UE} \ac{Rx} must properly steer the phased array beam towards the \ac{BS} \ac{Tx}. In current \ac{mmW} systems, this directional beam selection method, commonly known as \ac{BA} or beam training, requires synchronization signal correlation measurements for an exhaustive search over all possible \ac{Rx} beam directions. Since antenna arrays with more elements generate narrower directional beams, \ac{BA} communications overhead increases linearly with the array size. As future wireless systems at \ac{mmW} and \ac{sub-THz} frequencies require larger arrays for higher gain, exhaustive search \ac{BA} may be impractical.

Accelerated \ac{BA} is an active area of research, with prior work primarily using \ac{CS} or machine learning algorithms verified through simulations. Other solutions to \ac{BA} include hierarchical searches \cite{Hassanieh2018_AgileLink} \cite{Noh2017_hierarchical}; algorithms utilizing out-of-band information, such as node locations \cite{Heng2019_gpsBA} or sub-6 GHz channel information \cite{Alrabeiah2020_sub6}; and designs requiring novel array architectures \cite{Boljanovic2020_ttd}.

Several existing solutions apply \ac{CS} methods to solve \ac{BA}. \cite{Yan2019_CS} utilizes a \ac{MP} algorithm with channel gain measurements from \ac{PN} beams, quasi-omni-directional beams with random phase \ac{AWV}s. \cite{Rasekh2017_CSRSSMP} also employs \ac{MP} with \ac{PN} sounding beams, but only requires \ac{RSS} power measurements by solving phase-less \ac{BA} as a \ac{CPR} problem. SBG-Code from \cite{Li2019_SBGcode_PhaseCode} solves \ac{BA} for hybrid or digital arrays as a \ac{CPR} problem.

Machine learning methods have also been proposed for multipath or phase-less \ac{BA}. \cite{Polese2020_deepbeam} and \cite{Ma2020_cnnSampMultipath} each study \textit{sample-based} \ac{CNN} algorithms in multipath channels. As described later, sample-based \ac{IA} may not be practical for high \ac{mmW} \ac{IA}. mmRAPID \cite{Yan2020_mmrapid} and DeepIA \cite{Cousik2020_DeepIA} each only require phase-less \ac{RSS} measurements from \ac{PN} beams and pencil beams respectively, but do not thoroughly study the impact of multipath channels on the \ac{BA} performance. Thus, none of these works address phase-less \ac{mmW} \ac{BA} performance over varied multipath strength.

Few prior works have experimentally validated their \ac{BA} algorithms. \cite{Yan2020_mmrapid} used a 60 GHz testbed but with only single path and \ac{LOS} channels, while \cite{Ghasempour2019_MUTE} experimentally studied multi-user beam steering at 60 GHz but did not evaluate machine learning based \ac{BA}. \cite{Polese2020_deepbeam} tested a sample-based \ac{CNN} algorithm in a simple 60 GHz \ac{NLOS} channel, but did not thoroughly investigate the impact of multipath strength. To the authors' best knowledge, this work is the first to experimentally verify a machine learning based phase-less \ac{BA} algorithm at \ac{mmW} in multipath channels.

In this work, we propose an alternative algorithm for \ac{BA} to reduce the communications overhead. We develop our system model and problem statement in Section \ref{sec:system}, then describe our proposed phase-less \ac{BA} algorithm, including the foundation for this work in \cite{Yan2020_mmrapid}, in Section \ref{sec:algo_design}. In Section \ref{sec:eval_design}, we discuss our evaluation of the algorithms, including 60 GHz experiments, simulations, and tradeoff results. Finally, we conclude in Section \ref{sec:conclusion}.



\textit{Notation}: Scalars, vectors, and matrices are represented by non-bold lowercase, bold lowercase, and bold uppercase letters respectively. The $i$th column of a matrix $\mathbf{A}$ is denoted as $[\mathbf{A}]_i$. The transpose and Hermitian transpose of $\mathbf{A}$ are $\mathbf{A}^T$ and $\mathbf{A}^H$ respectively. $|\mathbf{A}|$ denotes the entry-wise magnitude of $\mathbf{A}$. 


\section{System Model and Problem Statement}
\label{sec:system}
This section formally defines our system model, the phase-less \ac{BA} problem, and limitations of existing solutions.

\subsection{Multipath Channel Model}
\label{subsec:channel}
For our algorithm design and simulations, we consider a simplified \ac{mmW} multipath channel model to provide a succinct metric for the relative strength of the \ac{NLOS} paths. To simplify the path gains for the few significant \ac{AoA}, we use a geometrically decaying model where each distinct path is successively attenuated by $e^{-\alpha}$. The model parameter $\alpha$ describes the channel gain and the \ac{BS} antenna response at the \ac{AoD} for each path. The channel is explicitly described in \eqref{eq:channel}, where $[\mathrm{\mathbf{a}_R}\left(\phi_l\right)]_n = a_n \exp{\left(j2\pi(n-1)\sin{(\phi_l)}d/\lambda\right)}$ is the $n$th element of the \ac{UE} \ac{Rx} array response and $a_n$ is the complex coefficient representing element-wise hardware mismatch. Ideally, $a_n = 1$ for every element, but this is difficult to achieve in practice. Due to physical channel characteristics, even multipath and \ac{NLOS} channels at \ac{mmW} typically only feature a few strong paths \cite{Akdeniz2014_mmWchannel}. In this work, we assume $L=3$ for simulations and design experiments with $L \in \{2, 3\}$.
\begin{align}
    \mathbf{h} = \textstyle\sum_{l = 1}^L e^{-\alpha l} 
    \mathrm{\mathbf{a}_R}\left(\phi_l\right), \quad \mathbf{h} \in \mathbb{C}^{N_r}
    \label{eq:channel}
\end{align}

\subsection{Problem Statement: Phase-less Beam Alignment}
\label{subsec:BA}
This work addresses user \ac{BA} with phase-less power measurements from a single, predefined sensing codebook. Phase-less measurements are specifically considered because devices generally have poor phase estimation during the early stages of \ac{IA}. Commodity hardware, like the 60 GHz radios used in this work, do not support phase measurements during \ac{BA}. Furthermore, the phase noise in hardware at upper-\ac{mmW} and \ac{sub-THz} frequencies would make phase information during \ac{BA} unreliable \cite{Bicais2019_phaseNoise}. We avoid algorithms with adaptive sensing codebooks (designs that change codebooks based on prior measurements, e.g. hierarchical searches) or out-of-band information to reduce the beam management circuit complexity.

Given a \ac{UE} with $N_r$ antennas, the sensing codebook with $M$ measurements can be represented as $\mathbf{W}_s \in \mathbb{C}^{N_r \times M}$, while the codebook used for data communications with $K$ directional beams  is $\mathbf{W}_d \in \mathbb{C}^{N_r \times K}$. The \ac{BS} is assumed to be transmitting a pilot sequence, either omni-directionally or with a beam aligned to the user before \ac{UE} \ac{BA}. The phase-less measurements for a given channel $\mathbf{h}$ is then shown in \eqref{eq:rss}.
\begin{align}
    \mathbf{y} = \left|\mathbf{W}_s^H\mathbf{h}s + \mathbf{W}_s^H\mathbf{n}\right| ,\quad \mathbf{y} \in \mathbb{R}^M
    \label{eq:rss}
\end{align}

The goal of \ac{BA} is ultimately to find the best directional beam for data communication, as in $i = \argmax_i \left|[\mathbf{W}_d]_i^H\mathbf{h}s\right|$. An estimated solution to this problem can be written as the objective for the traditional exhaustive search algorithm \eqref{eq:BAgeneral}. 
\begin{align}
    \label{eq:BAexhaustive}
    y_{d, i} &= \left|[\mathbf{W}_d]_i^H\mathbf{h}s + 
    [\mathbf{W}_d]_i^H\mathbf{n}\right| \\
    \label{eq:BAgeneral}
    \hat{\imath} &= \textstyle\argmax_i y_{d, i}^2
\end{align}

For algorithms using a non-adaptive sensing codebook, phase-less \ac{BA}'s objective is to estimate $\hat{\imath}$ using $\mathbf{y}$, as in \eqref{eq:BAcompressive}. Since the true probability is difficult to estimate, compressive \ac{BA} approximates $\tilde{\imath}$ with \ac{BA} algorithm $p(\mathbf{y})$. On the other hand in an exhaustive search, $p(\mathbf{y}) = \argmax_i y_{i}^2$ and $\mathbf{W}_s = \mathbf{W}_d$.
\begin{align}
    \label{eq:BAcompressive}
    \tilde{\imath} = \textstyle\argmax_i P(i = \hat{\imath} \mid \mathbf{y}) \approx p(\mathbf{y})
\end{align}



When designing $p(\mathbf{y})$, accuracy, gain loss, and number of required measurements serve as performance metrics. Since \eqref{eq:BAcompressive} represents a classification problem, the \textit{accuracy} of $p(\mathbf{y})$ for $N$ test points is the fraction of test points with the correct predicted beam direction: $acc(\hat{\imath},  p(\mathbf{y})) = \frac{1}{N}\sum_{j = 1}^N\mathbbm{1}[\hat{\imath} = p(\mathbf{y})]$. Accuracy, however, does not capture the impact of incorrect prediction on the data communication phase. In this work, the alignment quality is measured using maximum \textit{gain loss} $g = \mathbb{E}[\left|y_{d, \hat{\imath}}\right|^2]/\mathbb{E}[\left|y_{d,p(\mathbf{y})}\right|^2]$, or the difference in gain (in dB) between the optimal pencil beam and the pencil beam selected by the \ac{BA} algorithm, for a specified percentile of $N$ predictions. To evaluate the effective reduction in overhead, this work uses the \textit{required number of measurements} to meet a maximum gain loss requirement. 

The goal of this work is thus to demonstrate a machine learning based $p(\mathbf{y})$ and alternative $\mathbf{W}_s$ that reduce the required number of measurements for \ac{BA} in multipath channels. 


\subsection{Model-based Solutions and their Limitations}
\label{subsec:modelbased}
Phase-less \ac{CS} algorithms commonly use \ac{MP} with \ac{RSS} measurements (denoted \ac{RSS}-\ac{MP}) from specially designed sounding beams as $p(\mathbf{y})$. \cite{Rasekh2017_CSRSSMP} applies this method with quasi-omni-directional \ac{PN} beams. \ac{PN} beams use random phases for each antenna element, e.g. 2-bit phase \ac{AWV} $[\mathbf{W}_s^{PN}]_i$ with $n$th entry $w_n = \exp{(\phi_{n})}, \phi_{n} \in \{0, \frac{\pi}{2}, \pi, \frac{3\pi}{2}\}$, creating random antenna patterns with low angular correlation between beams. Alternatively, the adaptive codebook and scoring system in \cite{Hassanieh2018_AgileLink} can be reformulated as an \ac{RSS}-\ac{MP} algorithm with multifinger beams with multiple directional lobes. In this work, these two \ac{MP} algorithms are used as baselines for comparison.


Practical hardware and multipath channels provide challenges for phase-less, model-based \ac{BA} performance. \ac{RSS}-\ac{MP} with \ac{PN} beams rely on perfect prior knowledge of the dictionary, $\mathbf{W}_s^{PN}$, for good spatial reconstruction. \cite{Yan2020_mmrapid} experimentally demonstrated that \ac{PN} beam \ac{RSS}-\ac{MP} is vulnerable to poor array calibration and hardware impairments. Additionally, signals from different channel paths can interfere destructively in  phase-less, compressive measurements, compromising angular path distinction. In the noise-free case below, interference between different paths are included in the second summation, distorting the \ac{RSS} measurement $\left|y_i\right|^2$.
\begin{align*}
    \left|y_i\right|^2 
    & = \quad \left|
    \left[\mathbf{W}_s\right]_i^H\mathbf{h}
    \right|^2 \\
     & = \quad \textstyle\sum_{l}
    |\alpha_l|^2
    \left|\left[\mathbf{W}_s\right]^H_i
    \mathbf{a_R}(\phi_{l_1})\right|^2 + \\
    &\textstyle\sum_{l_1\neq l_2}
    2\mathcal{R}e\left\{
    \alpha_{l_1}^H
    \alpha_{l_2}
    \mathbf{a_R}(\phi_{l_1})^H\left[\mathbf{W}_s\right]_i
    \left[\mathbf{W}_s\right]^H_i
    \mathbf{a_R}(\phi_{l_2})
    \right\}
\end{align*}
 
To reduce magnitude of the second term, \textit{sparsely supported} beams are necessary. Beams, $\left[\mathbf{W}_s\right]_i$, that are sparsely supported in the angular space
only have large gain for few angles $\phi_l$, reducing the probability of interference with other \ac{AoA}s. \cite{Li2019_SBGcode_PhaseCode} demonstrated this property with theoretical bounds on \ac{BA} performance, while \cite{Hassanieh2018_AgileLink} utilized this principle with phased array multifinger beams. A simple numerical demonstration of the need for sparsely supported beams is given below. Suppose the channel $\mathbf{h} = \mathbf{W}_d^H \hat{\mathbf{x}}$ is on-grid for vector $\hat{\mathbf{x}} \in \mathbb{C}^{K}, \lVert\hat{\mathbf{x}}\rVert_0 = L$ representing a sparse set of paths with largest entry at index $\hat{\imath}$. Consider the following model-based algorithm that estimates $|\hat{\mathbf{x}}|$ by solving a phase-less version of \ac{MP} \eqref{eq:sparseNumerical}.
\begin{align}
    \label{eq:sparseNumerical}
    \min_{\mathbf{x},\epsilon} & \quad \gamma \left\|\mathbf{x}\right\|_1 + \epsilon  \\
    \mathrm{s.t.} & \quad 
    \left\|\left|\mathbf{W^H_sW_d}\right|\mathbf{x} - |\mathbf{y}|\right\|_2^2 \leq \epsilon\|\mathbf{y}\|_2^2, \nonumber\\
    & \quad \mathbf{x}\succeq 0 \nonumber
\end{align}

\begin{figure}[htbp!]
    \vspace{-3mm}
    \centering
    \includegraphics[width = 1.0\linewidth,trim=20 0 0 2,clip]{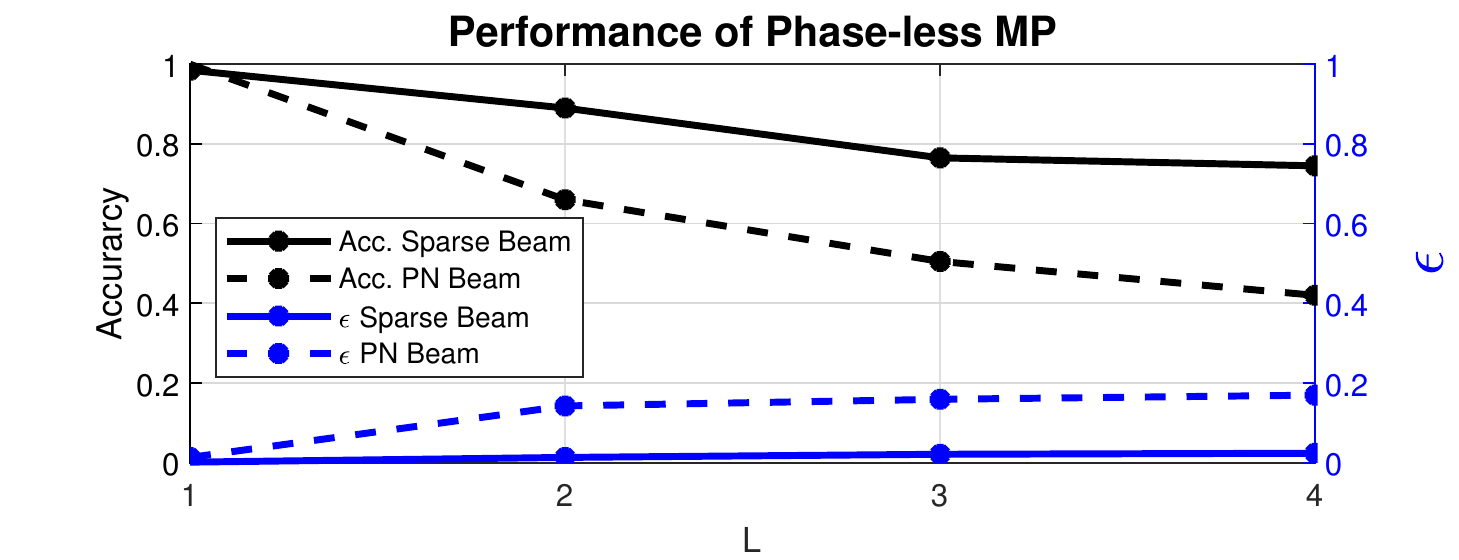}
    \vspace{-6mm}
    \caption{\ac{MP}
    performance vs. number of channel paths}
    \label{fig:23}
    \vspace{-4mm}
\end{figure}

Here, $\mathbf{x}$ is an estimate of $|\hat{\mathbf{x}}|$, $\epsilon$ captures distortion in $\mathbf{y}$ due to interference between multipath components, and $\gamma$ is a trade-off constant for the tolerance to $\epsilon$. For each realization of the channel, this algorithm's performance is evaluated with two codebooks $\mathbf{W}_s$: PN beams $\mathbf{W}_s^{PN}$ and one containing sparse beams encoding all $K$ on-grid angles from \cite{Li2019_SBGcode_PhaseCode}. 
Fig. \ref{fig:23} shows the simulated accuracy, with $N_r = K = 72, M = \frac{N_r}{3} = 24, \gamma = 0.01$, under different channel sparsity $L \in \{1,2,3,4\}$. To ensure convergence of $\epsilon$, $ |\mathbf{y}|$ is normalized by its maximum entry before using the CVX solver. The accuracy, $acc(\hat{\imath}, \tilde{\imath})$, compares  $\hat{\imath}$ with predictions $\tilde{\imath} = \argmax_i |\mathbf{x}|_i$. The \ac{PN} codebook achieves a high accuracy in predicting \ac{AoA} with no multipath components ($L = 1$), but degrades drastically with $L > 1$, requiring a relaxed $\epsilon$ to get a feasible solution. Performance of the sparse beams remains high even with $L = 4$ and an order of magnitude smaller $\epsilon$, demonstrating that sparse beams are more consistent in multipath channels.
 
\section{Algorithm Design}
\label{sec:algo_design}
To reduce \ac{BA} overhead in multipath channels, this paper proposes an improved version of mmRAPID \cite{Yan2020_mmrapid}. Like mmRAPID, the approach uses the same two-stage lifecycle and classification formulation to reduce the impact of hardware impairments. However, the proposed algorithm utilizes a novel combination of sounding beams and a \ac{CNN} architecture to exploit correlation between the new features. This section provides insight into these design choices and tradeoffs.

\subsection{Two-Stage Data-Driven Phase-less Beam Alignment}
\label{subsec:ml2stage}
The two-stage \ac{ML} algorithm design in mmRAPID serves as the foundation for this work. During the training stage, the \ac{UE} conducts both an exhaustive search using the directional codebook $\mathbf{W}_d$ and the compressive measurements with the sensing codebook $\mathbf{W}_s$. After collecting sufficient data, the machine learning classification model is trained using the best directions as labels and the sensing measurements as features. In the \ac{ML} architectures, this translates to an input layer of size $M$ and an output score vector of size $K$. During the testing stage, the \ac{UE} only uses $\mathbf{W}_s$ for \ac{BA} sounding. Although the \ac{UE} will see higher \ac{IA} overhead during the training stage, the overhead is much smaller during the testing stage and the majority of device usage. 

\subsection{Sounding Beam Design}
\label{subsec:beams}
This work considers three heuristic beam designs for the sounding codebook $\mathbf{W}_s$: \ac{PN} beams, \ac{SA} multifinger beams, and single lobe \ac{QPD} beams. The \ac{PN} beams in this work are defined with 2-bit phase (see Section \ref{subsec:modelbased}). As shown in Fig.\ref{fig:23}, \ac{PN} beams are not ideal for phase-less multipath \ac{BA}, thus the sparser multifinger beams and \ac{QPD} beams are tested.


\ac{SA} multifinger beams, abbreviated as \ac{SA} beams, have multiple directional lobes with lower maximum gain than full pencil beams. As in \cite{Hassanieh2018_AgileLink}, \ac{SA} beam \ac{AWV}s are generated by concatenating weight vectors for independently-steered pencil beams from single subsections of the phased array. Formally, the \ac{AWV} for the $i$th sub-array, pointing toward $\theta_i$, is $w_{i,n} = \exp{(\phi_{1,n})}, \phi_{i,n}=2\pi n\sin(\theta_i) d/\lambda$. 


\ac{QPD} beams from \cite{Sayidmarie2013_QPD} use phase adjustments to widen pencil beams using the entire phased array. Increasing design parameter $\Phi > 0$ increases the added phase $\phi_{n,qpd} = 4 \Phi \times \left(\tfrac{2n-(N+1)}{2(N+1)}\right)^2$ and the beamwidth, but decreases the maximum gain. The overall \ac{AWV} for a \ac{QPD} beam pointing toward $\theta$ is $[\mathbf{W}_s^{QPD}]_{i,n} = \exp{(\phi_{n})}, \phi_{n}=2\pi n\sin(\theta) d/\lambda + \phi_{n,qpd}$. This work assumes $\Phi=\pi$ for sparsely supported beams with the same beamwidth as an \ac{SA} beam finger, but with higher gain.

 
\begin{figure}
    \centering
    \vspace{1mm}
    \includegraphics[width=0.47\textwidth,trim=120 120 120 120,clip]{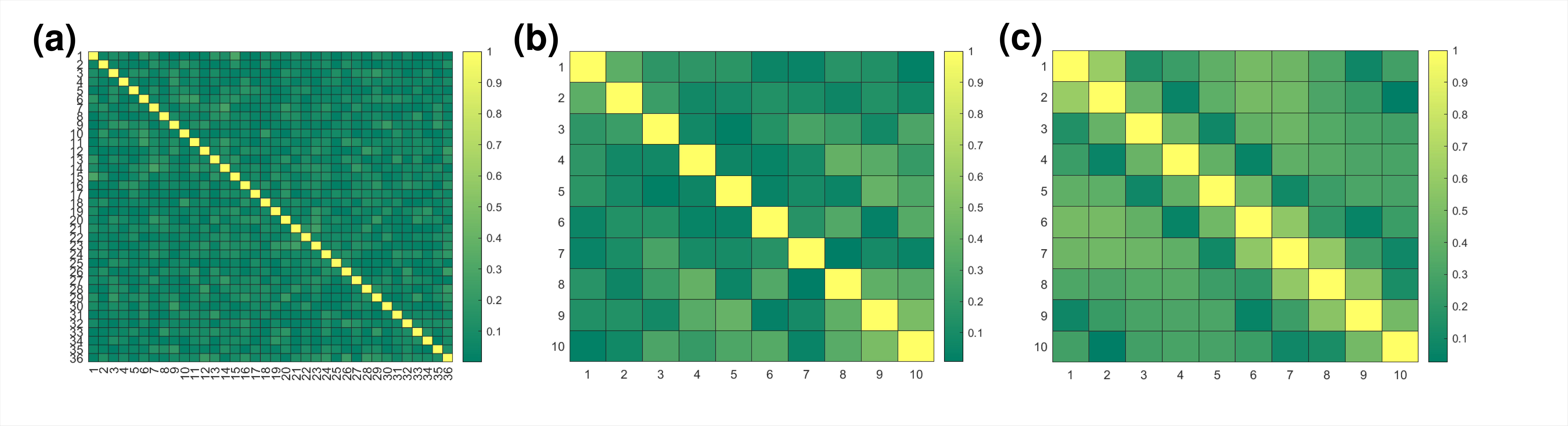}
    \caption{Correlation between \ac{RSS} features in a multipath channel. (a) \ac{PN} beams; (b) \ac{SA} beams; (c) \ac{QPD} beams}
    \label{fig:correlation}
    \vspace{-7mm}
\end{figure}

\subsection{Machine Learning Architecture}
\label{subsec:MLarch}
While the mmRAPID algorithm used a \ac{MLP} for \ac{ML} based beam alignment, alternative algorithms may provide better \ac{AoA} prediction. This work investigates \ac{CNN}s to improve phase-less \ac{BA} over \ac{MLP}s. \ac{CNN}s are commonly used in image processing for their ability to utilize correlation in the feature space. Thus, if the \ac{RSS} measurements from the sounding beams are correlated for a given channel, a \ac{CNN} may improve performance. Data collected from the 60 GHz testbed, described in Section \ref{subsec:exp_design}, show significant correlation in the feature space for \ac{SA} and \ac{QPD} beams. Fig. \ref{fig:correlation} shows the average experimental correlation found between features in a 2000 realizations of a multipath channel with $L=2$ and $\alpha ~= 0.2$. This correlation encourages the study of \ac{CNN}s as a potential prediction algorithm.

\begin{figure}
    \centering
    \includegraphics[width=0.47\textwidth,trim=100 110 100 100,clip]{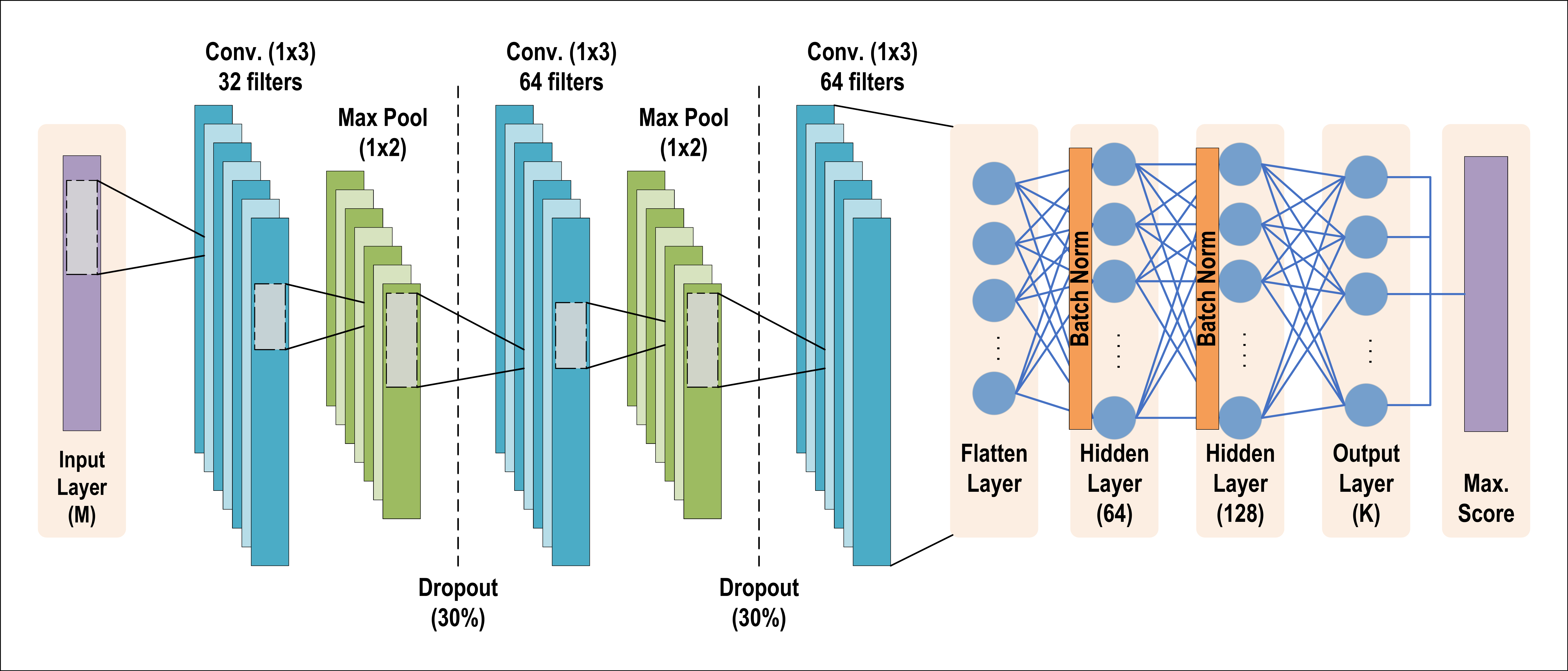}
    \caption{\ac{CNN} architecture considered}
    \label{fig:cnnArch}
    \vspace{-1mm}
\end{figure}
Fig. \ref{fig:cnnArch} details the tuned \ac{CNN} architecture. The convolutional layers extract features for the following \ac{MLP}. Note that the evaluated \ac{MLP} prediction model is the same as the \ac{CNN}'s \ac{MLP} stage, other than an input layer replacing the flatten layer. Both use ReLU activation functions for the outputs of layers and are trained with the RMSprop optimizer. As the algorithms aim to solve \ac{BA} as a classification problem, both designs use sparse categorical cross entropy as a loss function. Both the \ac{MLP} and \ac{CNN} algorithms were developed in Tensorflow using the Keras API, with network hyperparameters selected using separate validation data from past experiments. \textit{The code and dataset will be available upon publication.}

\section{Algorithm Evaluation}
\label{sec:eval_design}


\subsection{Experimental Testbed}
\label{subsec:exp_design}  
\begin{figure}
    \vspace{-3mm}
    \begin{center}
    \includegraphics[width=0.45\textwidth,trim=20 20 20 20,clip]{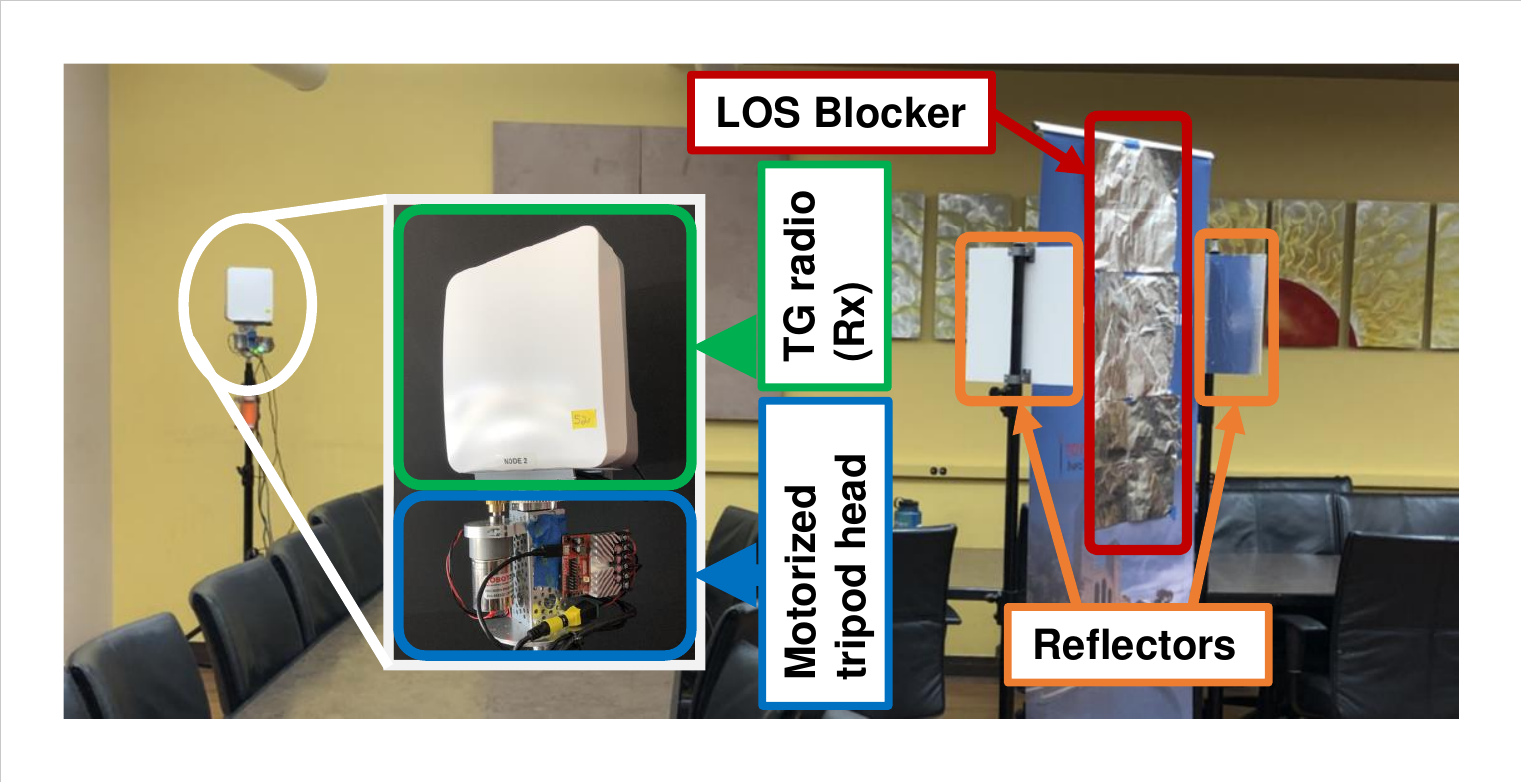}
    \end{center}
    \vspace{-3mm}
    \caption{Testbed \ac{Rx} with two reflectors and a \ac{LOS} blockage.}
    \vspace{-6.5mm}
    \label{fig:testbed}
\end{figure}
To capture the impact of practical hardware impairment, the algorithms were verified using a 60 GHz testbed featuring two Facebook \ac{TG} radios. Using IEEE 802.11ad packet preambles, the \ac{TG} radios measure received power and SNR estimates. Each \ac{TG} radio features a $36 \times 8$ element phased array, with programmable phase in the 36 element horizontal axis. The programmable phase enabled measurements with custom azimuthal beamforming used in the sounding codebook design discussed in Section \ref{subsec:beams}. 

The receiver TG radio was attached to a programmable motorized turntable, enabling the automated data collection required for the \ac{ML} datasets. For a given channel environment with manually-placed reflectors, obstructions, and \ac{TG} nodes, the software controls the radios and turntable to capture channel measurements with varied physical angles between the two phased arrays. Each new angle between the \ac{Tx} and \ac{Rx} geometrically provides a new channel configuration. Fig.~\ref{fig:testbed} shows the \ac{Rx} TG radio with the motorized turntable. 

\subsection{Simulation Design and Data Preparation}
\label{subsec:sim_design}
We created two separate datasets based on simulations and measurements. The simulations study a wider variety of multipath channels, while the experiments investigate hardware impairment impact. Even with automated collection, the number of experimental channels was limited by time constraints. The experimental data includes 8 $\alpha$ configurations, each with fixed relative separation between \ac{AoA}s, rotated to 2000 different physical \ac{AoA}s. However, the simulations include 9 $\alpha$'s, each with 10 measurements from 400 \ac{AoA} relative separations, totaling 4000 channel realizations per $\alpha$. Simulated paths were randomly selected over a $30^{\circ}$ range, only requiring at least $2^{\circ}$ of separation to maintain distinct \ac{AoA}s.


Both simulated and experimental data are validated and split into separate training and testing sets. Each label used 144 samples for training, with the remaining data used for testing. With these fixed training set sizes, the simulated \ac{ML} algorithms train with fewer examples of each \ac{AoA} combination. Data validation removed points with labels without the required 144 training points. In this work, validation left $K=54$ and $K=51$ pencil beam labels (and maximum exhaustive search \ac{BA} overhead) for simulations and experiments respectively.


\subsection{Evaluation Results}
\label{subsec:exp_results}
\begin{figure}
     \centering
     \begin{subfigure}[c]{0.47\textwidth}
         \centering
         \includegraphics[width=\textwidth,trim=10 0 20 0,clip]{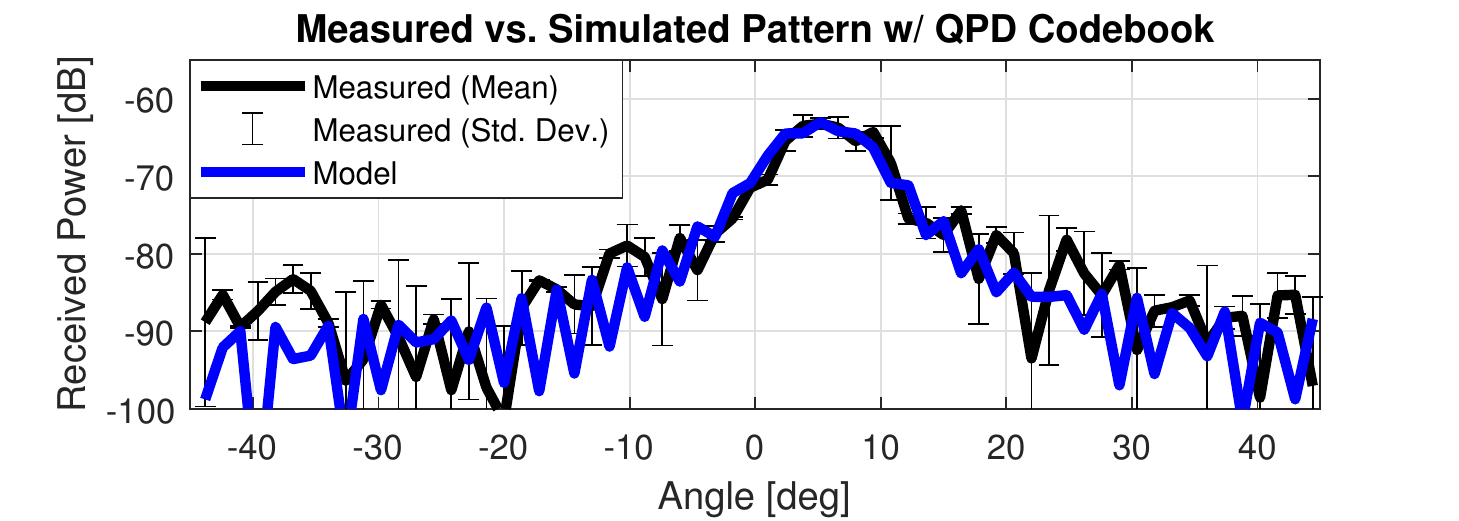}
         \vspace{-5mm}
         \caption{\ac{QPD} beam}
         \label{fig:beam_meas_qpd}
     \end{subfigure}
     \begin{subfigure}[c]{0.47\textwidth}
         \centering
         \includegraphics[width=\textwidth,trim=10 0 20 0,clip]{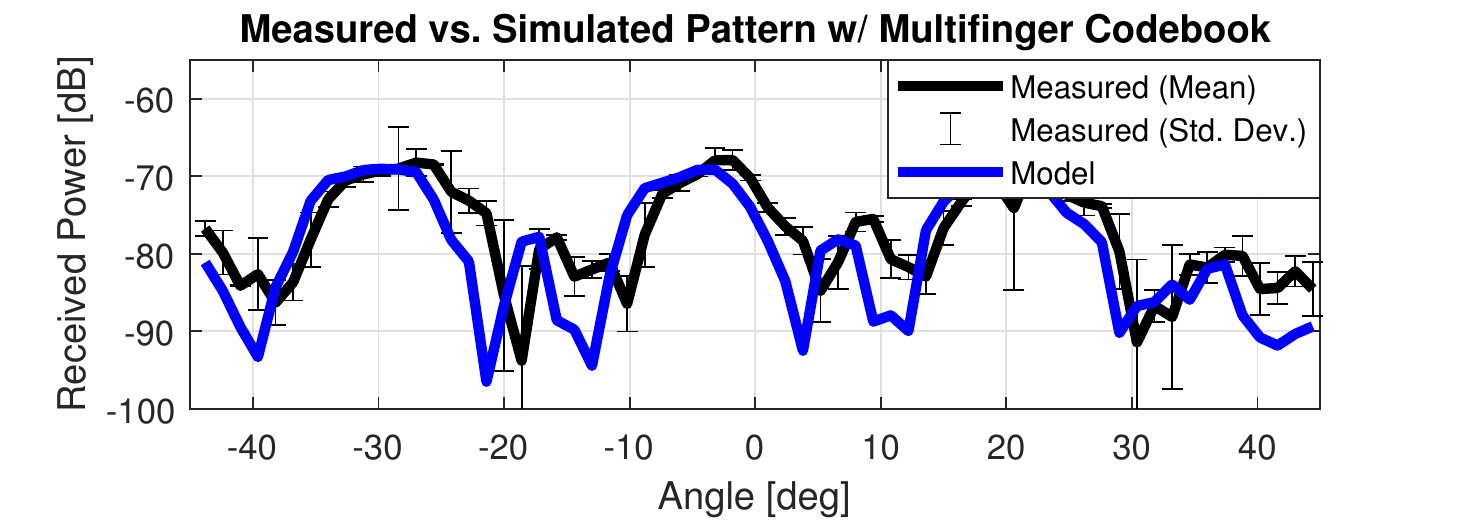}
         \vspace{-5mm}
         \caption{\ac{SA} multifinger beam}
         \label{fig:beam_meas_sa}
     \end{subfigure}
     \vspace{-1mm}
    \caption{Theoretical (blue) and measured (black) beam patterns.}
    \label{fig:beam_meas}
    \vspace{-7mm}
\end{figure}

This work empirically compares the discussed algorithms to find tradeoffs for $\mathbf{W}_s$ and $p(\mathbf{y})$  designs. Both simulations and experiments used the same $\mathbf{W}_d$ and $\mathbf{W}_s$, but no impairments were added in the simulations. $\mathbf{W}_d$ included 64 pencil beams over \ac{TG}'s supported angular range of $[-45^\circ, 45^\circ]$, oversampling the \ac{AoA}s. For baseline methods, $\mathbf{W}_s^{PN}$ included 36, 2-bit \ac{PN} beams. Each \ac{SA} beam used three, 12-element virtual sub-arrays with $25^{\circ}$ separation between fingers. Both \ac{SA} and \ac{QPD} codebooks include 10 beams with even angular separation ($9^{\circ}$) between the centers of codes. Fig. \ref{fig:beam_meas} presents the theoretical and experimentally measured beam patterns of the one \ac{QPD} and one \ac{SA} beam, highlighting the impact of hardware impairments on the true pattern. As with \ac{PN} beams in \cite{Yan2020_mmrapid}, these pattern offsets reduce \ac{MP} algorithm performance.  

\begin{figure}
     \centering
     \begin{subfigure}[c]{0.47\textwidth}
         \centering
         \includegraphics[width=\textwidth,trim=20 0 20 0,clip]{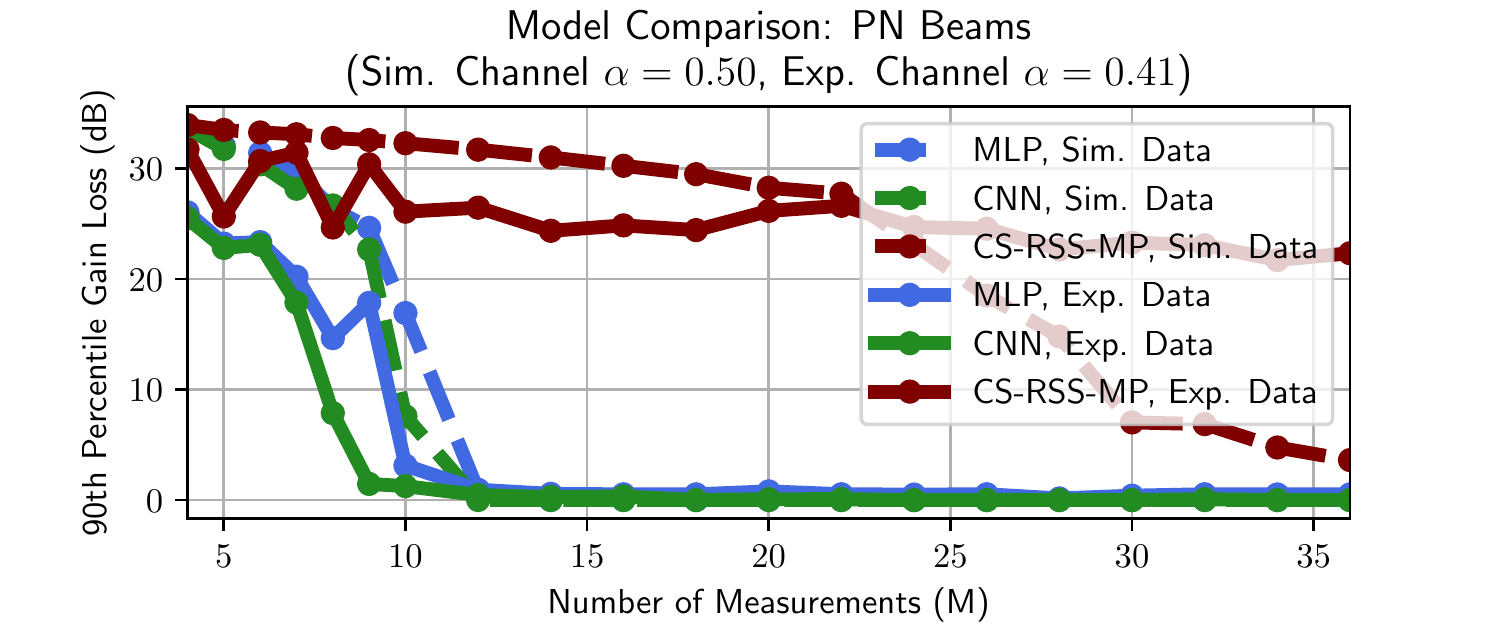}
         \vspace{-6mm}
         \caption{Algorithms using only \ac{PN} beams.}
         \label{fig:compModels_pn}
     \end{subfigure}
     \begin{subfigure}[c]{0.47\textwidth}
         \centering
         \includegraphics[width=\textwidth,trim=20 0 20 0,clip]{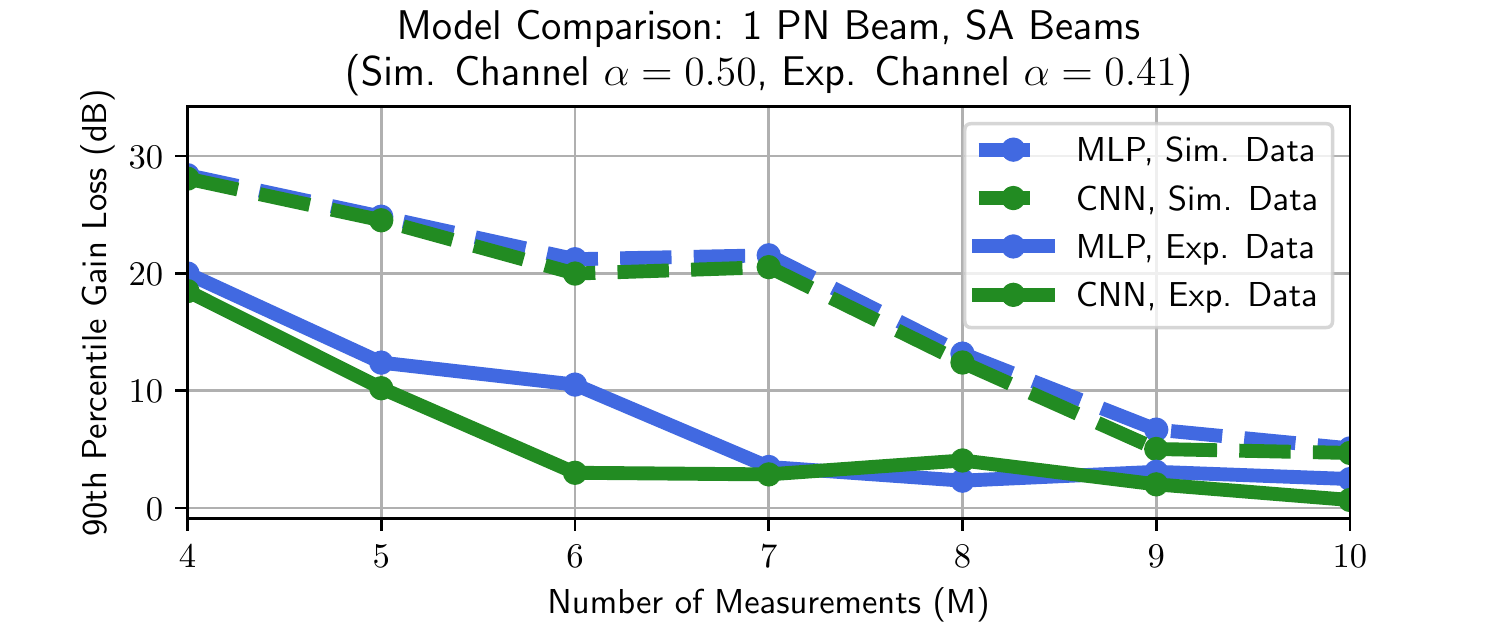}
         \vspace{-6mm}
         \caption{Algorithms using 1 \ac{PN} beam and the rest as \ac{SA} beams.}
         \label{fig:compModels_1pn-sa}
     \end{subfigure}
     \begin{subfigure}[c]{0.47\textwidth}
         \centering
         \includegraphics[width=\textwidth,trim=20 0 20 0,clip]{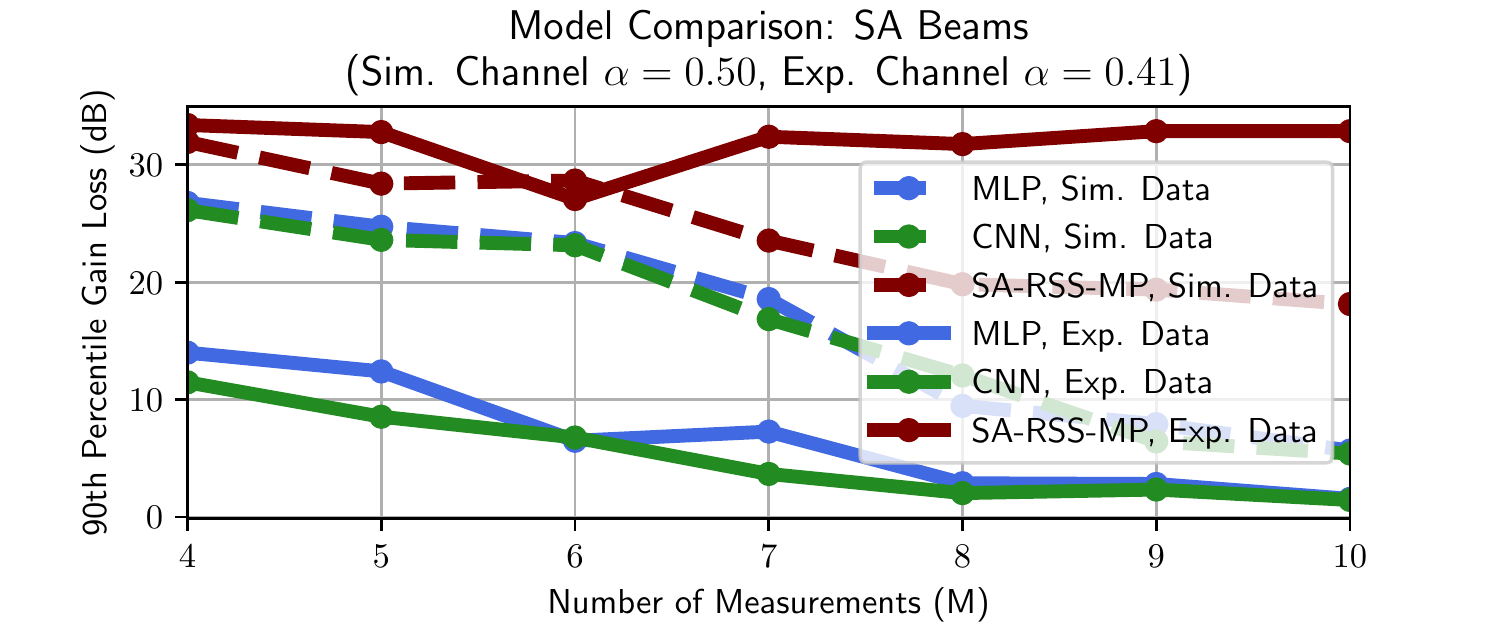}
         \vspace{-6mm}
         \caption{Algorithms using only \ac{SA} beams.}
         \label{fig:compModels_sa}
     \end{subfigure}
     \vspace{-1mm}
    \caption{Prediction algorithms under strong \ac{NLOS} \ac{AoA}s.}
    \label{fig:compModels}
    \vspace{-5mm}
\end{figure}

\begin{figure*}
    \centering
    \begin{subfigure}[b]{0.47\textwidth}
        \centering
        \includegraphics[width=\textwidth,trim=20 0 20 0,clip]{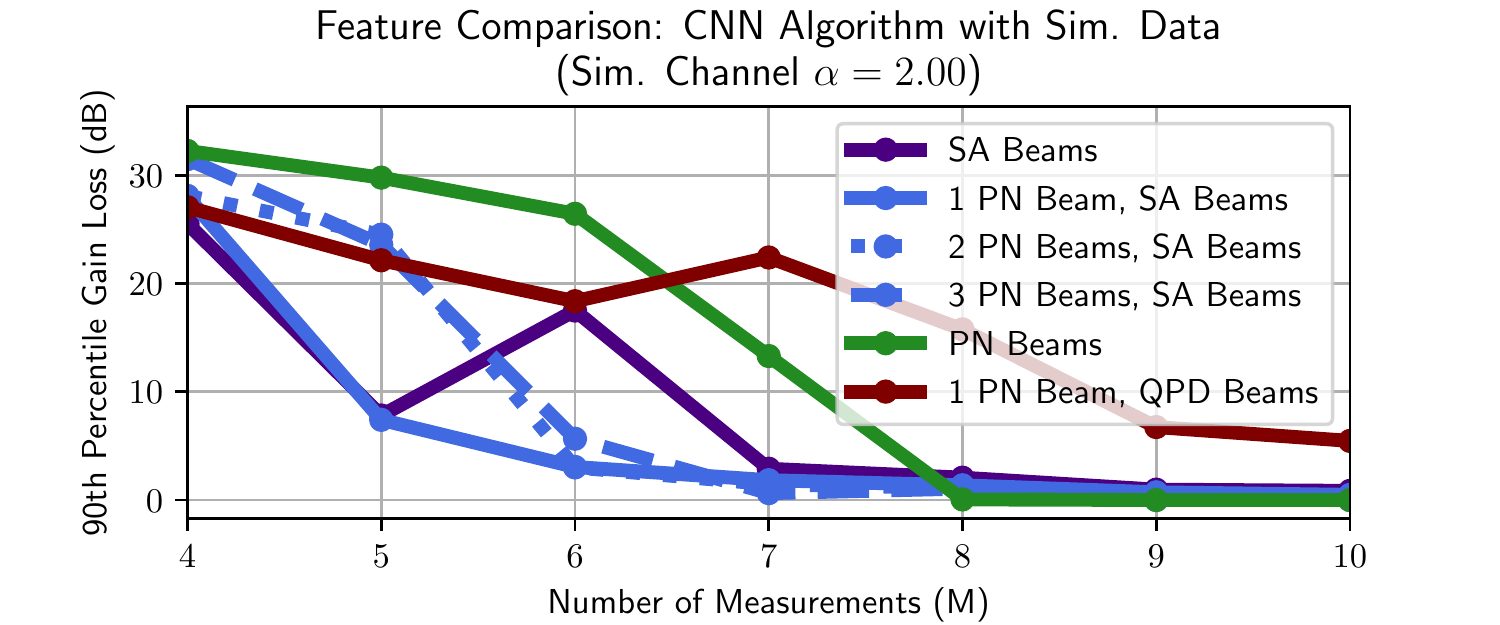}
        \vspace{-6mm}
        \caption{Simulation data, Weak \ac{NLOS}}
        \label{fig:compBeams_weak_sim}
    \end{subfigure}
    \hfill
    \begin{subfigure}[b]{0.47\textwidth}  
        \centering 
        \includegraphics[width=\textwidth,trim=20 0 20 0,clip]{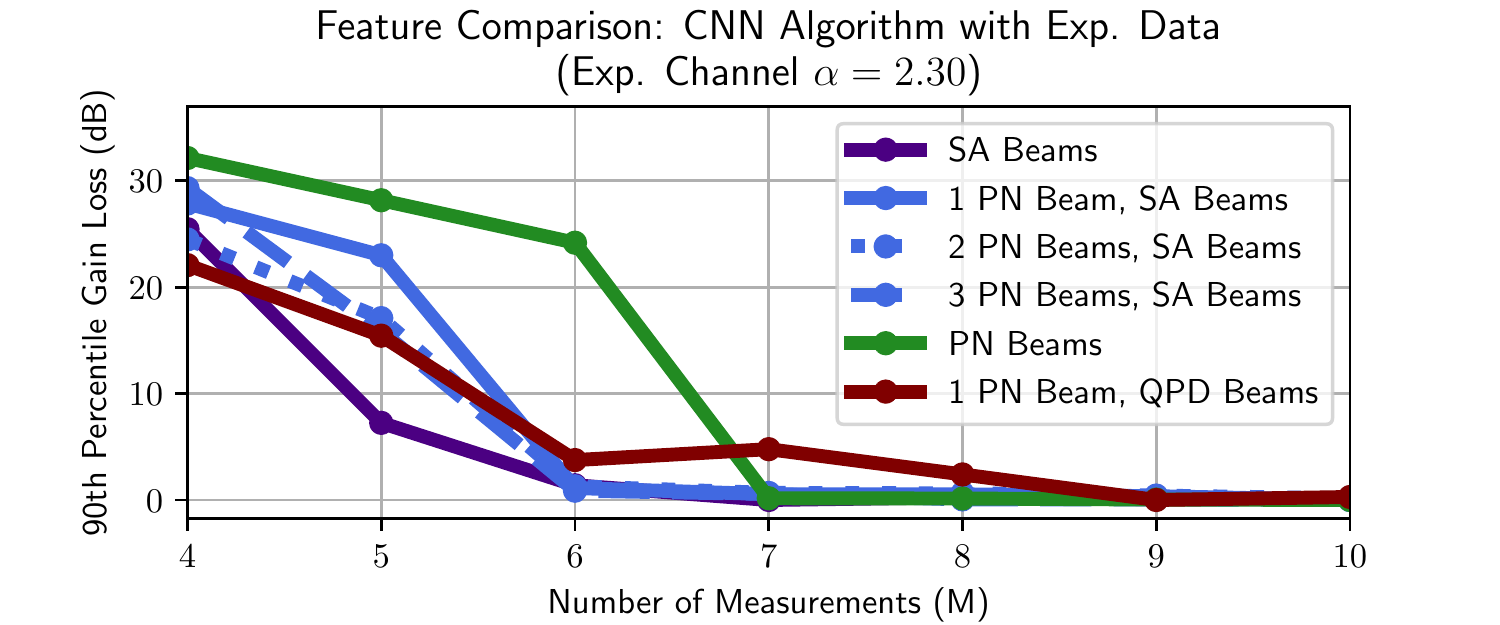}
        \vspace{-6mm}
        \caption{Experimental data, Weak \ac{NLOS}}    
        \label{fig:compBeams_weak_exp}
    \end{subfigure}
    \vskip\baselineskip
    \vspace{-4mm}
    \begin{subfigure}[b]{0.47\textwidth}   
        \centering 
        \includegraphics[width=\textwidth,trim=20 0 20 0,clip]{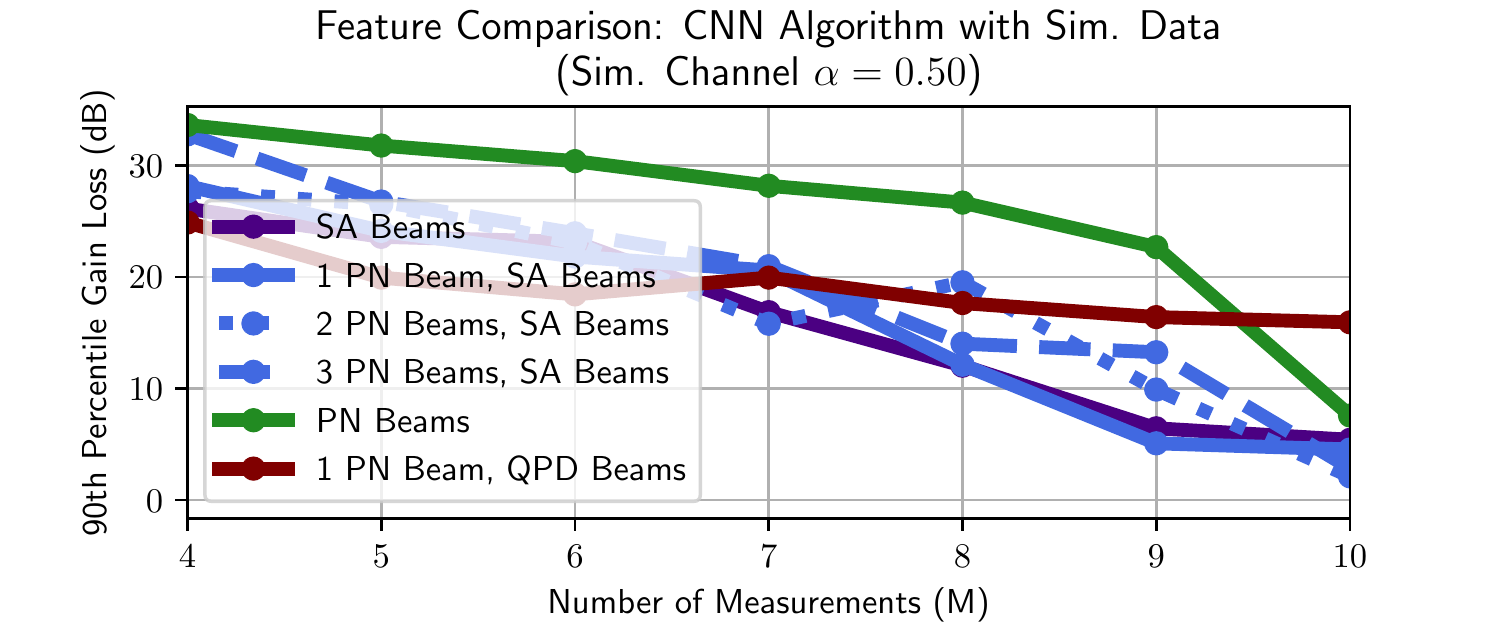}
        \vspace{-6mm}
        \caption{Simulation data, Strong \ac{NLOS}}   
        \label{fig:compBeams_med_sim}
    \end{subfigure}
    \hfill
    \begin{subfigure}[b]{0.47\textwidth}   
        \centering 
        \includegraphics[width=\textwidth,trim=20 0 20 0,clip]{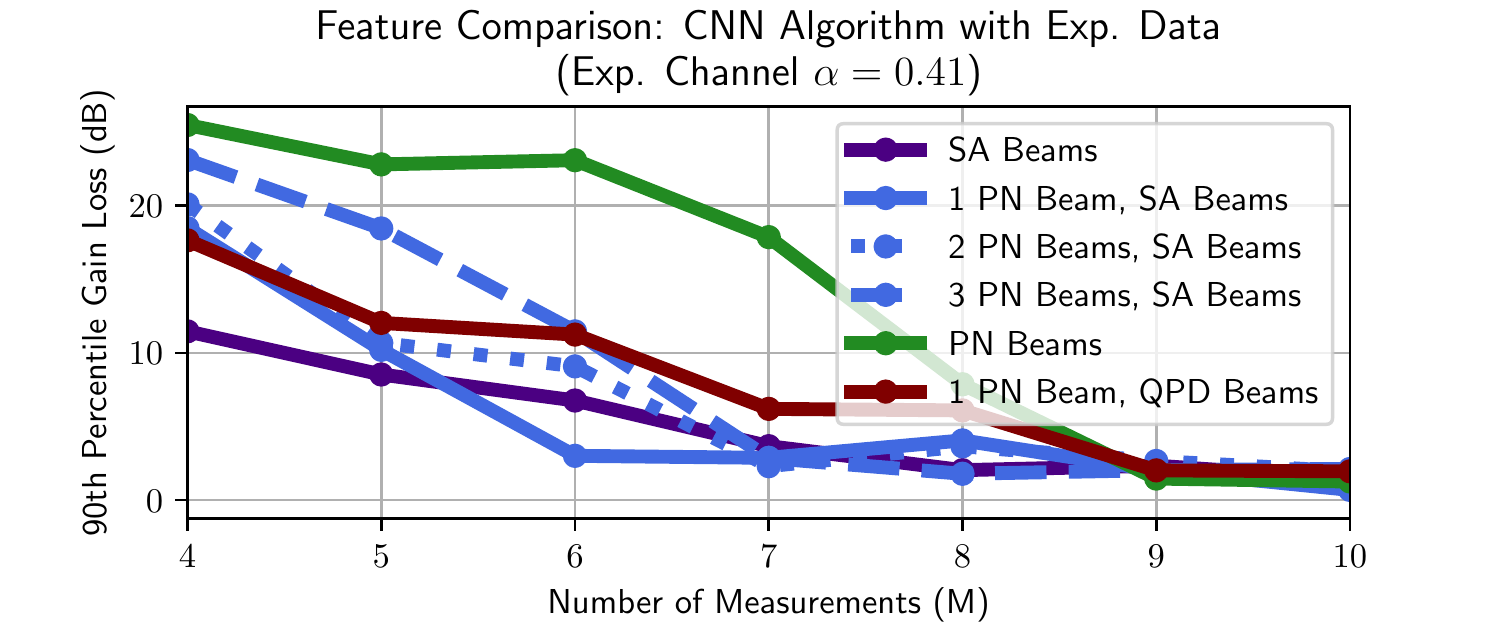}
        \vspace{-6mm}
        \caption{Experimental data, Strong \ac{NLOS}}  
        \label{fig:comp_beams_med_exp}
    \end{subfigure}
    \vspace{-1mm}
    \caption{Comparison of beam designs between simulations (Sim.) and experiments (Exp.) with strong and weak \ac{NLOS} \ac{AoA}s.} 
    \vspace{-5mm}
    \label{fig:compBeams}
\end{figure*}

\subsubsection{Beam Design Impact}
\label{subsubsec:beams}
The \ac{BA} gain loss of $\mathbf{W}_s$ designs is first compared with \ac{CNN} prediction algorithms and two channel configurations. Fig. \ref{fig:compBeams} provides this comparison for simulated and experimental results with weak and strong multipath components. As expected with wide angular support, exclusive use of \ac{PN} beam features provides nearly the worst performance in all four tests. While \ac{QPD} beams have much less angular support than \ac{PN} beams, this lack of angular coverage also provides less information to predict \ac{AoA}s and reduces \ac{BA} performance. In most cases, the lowest gain loss was reported for measurement combinations with \ac{SA} beams and a few \ac{PN} beams.

\subsubsection{Prediction Algorithm Comparison}
\label{subsubsec:models}
The prediction models were then compared to determine the best $p(\mathbf{y})$. Fig. \ref{fig:compModels} shows the comparison of \ac{CNN}, \ac{MLP}, and \ac{MP} with three combinations of features. \ac{MP} is used with the \ac{PN} and \ac{SA} codebooks to compare \ac{ML} algorithms with previously described model-based methods. In all cases, the \ac{CNN} algorithm provided the lowest gain loss, with order-of-magnitude reductions compared to \ac{MP} algorithms and improvements ranging from 0 to 20 dB compared to the \ac{MLP} algorithm. While the experimental and simulation results are superimposed, each dataset studies separate system parameters. Experiments show larger gain loss improvements in using \ac{ML} methods over \ac{MP} due to array impairments. Additionally, multipath forces the \ac{ML} algorithms to learn from both the impact of hardware impairment and the training data channels. Thus, experimental results, with fewer combinations of path \ac{AoA}s and gains, show better performance than simulations. 




\subsubsection{Required Number of Measurements}
\label{subsubsec:reqMeas}
\begin{figure}[t]
    \centering
    \begin{subfigure}[b]{0.49\columnwidth}
        \centering
        \includegraphics[width=\textwidth,trim=5 5 5 10,clip]{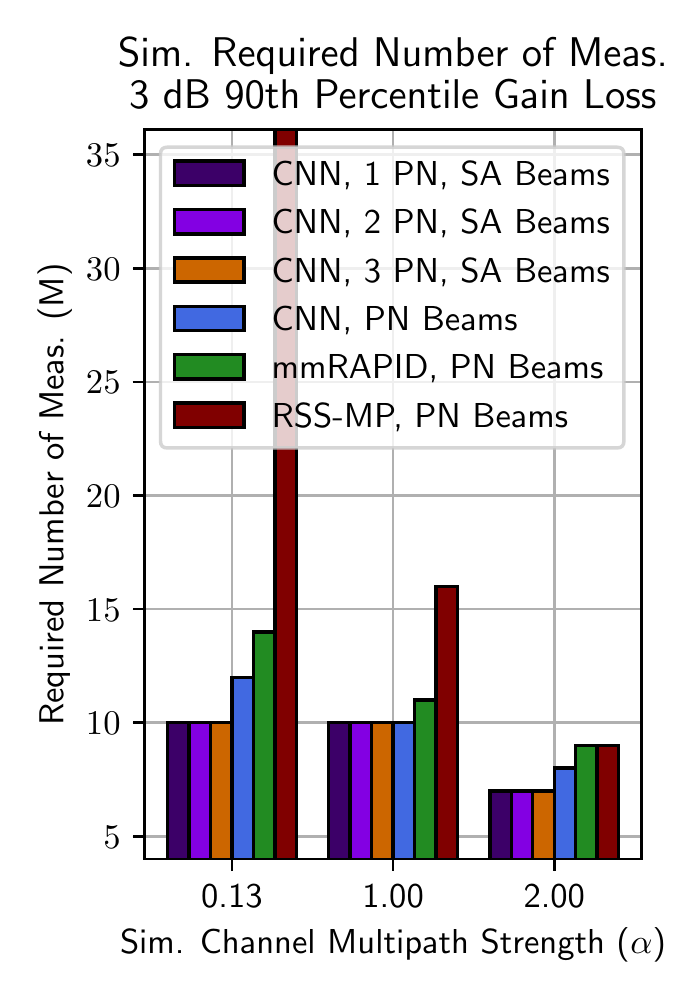}
        \vspace{-6mm}
        \caption{Simulated results}
        \label{fig:reqMeas_sim}
    \end{subfigure}
    \begin{subfigure}[b]{0.49\columnwidth}
        \centering
        \includegraphics[width=\textwidth,trim=5 5 5 10,clip]{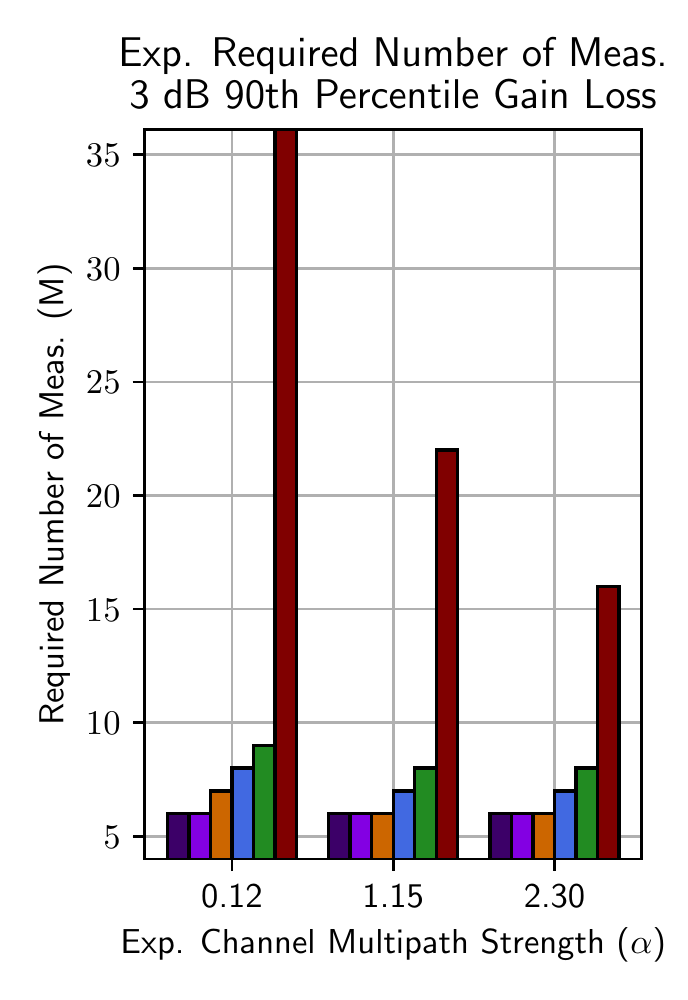}
        \vspace{-6mm}
        \caption{Experimental results}
        \label{fig:reqMeas_exp}
    \end{subfigure}
    \vspace{-1mm}
    \caption{Comparison of the required number of measurements to meet a 3 dB 90th percentile maximum required gain loss.}
    \label{fig:reqMeas}
    \vspace{-5mm}
\end{figure}

Based on the prior results, the best algorithm combination uses a combination of \ac{PN} and \ac{SA} beams for $\mathbf{W}_s$ and the \ac{CNN} for $p(\mathbf{y})$. Fig. \ref{fig:reqMeas} compares the required number of measurements for the proposed algorithm combinations, mmRAPID, and \ac{PN}-\ac{RSS}-\ac{MP}. Note that \ac{SA}-\ac{RSS}-\ac{MP} is not included, as the algorithm did not sufficiently reduce the gain loss with the 10 beams used. The best combination, the \ac{CNN} with 1 or 2 \ac{PN} beams and \ac{SA} beams, significantly decreases the \ac{BA} communications overhead. While an exhaustive search would require $K=51$ measurements in experiments, the best algorithm only requires 6, reducing overhead by 88\%. \ac{PN}-\ac{RSS}-\ac{MP} does not meet gain loss requirements with even 36 measurements in the strongest multipath channels and requires 16 in the best experimental case, demonstrating that the best algorithm reduces overhead by 63\% or more. Compared to mmRAPID's 8 or 9 required measurements in experiments, the best algorithm still reduces overhead by over 25\%. The proposed algorithm's benefits vary more in simulations, but still present significant performance improvements in all cases.

\section{Conclusion and Future Work}
\label{sec:conclusion}
This paper proposes a novel phase-less \ac{UE} \ac{BA} algorithm for multipath channels, using a mixed beam design codebook and a \ac{CNN} architecture. The design optimizes the heuristic codebook by empirically finding a balance between the sparse angular support required for phase-less prediction and the omnidirectionality required for a minimal number of measurements. The \ac{CNN} architecture takes advantage of the feature space correlation of these beam designs. Using  1 or 2 \ac{PN} beams and \ac{SA} beams significantly reduces the required number of measurements for effective \ac{BA}, demonstrating 88\% and at least 63\% less overhead in \ac{mmW} experiments as compared to an exhaustive search and \ac{MP} methods respectively.

The \ac{AoA} prediction method and beam design in this work warrant future improvement. Improving the algorithm to predict multiple best \ac{AoA}s could be helpful for backup connection beams in case of blockage. Additionally, a more comprehensive study of theoretically optimal sounding beam designs for phase-less \ac{BA} has not yet been explored.

\section*{Acknowledgment}
The \SI{60}{\giga\hertz} Terragraph channel sounders were gifts from the Telecom Infra Project (TIP).

\bibliographystyle{IEEEtran}
\bibliography{IEEEabrv,references}

\end{document}